\documentclass{aastex631}
\usepackage[flushleft]{threeparttable}
\usepackage{booktabs}
\usepackage{footmisc}
\usepackage{lineno}

\shorttitle{Toroid surrounding  W75N(B)-VLA2}
\shortauthors{G\'omez, J. F., et al.}

\begin{document}

\title{An SiO Toroid and Wide-angle Outflow associated with the Massive Protostar W75N(B)-VLA2}

\correspondingauthor{Jos\'e F. G\'omez}
\email{jfg@iaa.es}

\author[0000-0002-7065-542X]{Jos{\'e} F. G{\'o}mez}
\affiliation{Instituto de Astrof{\'i}sica de Andaluc{\'i}a (IAA-CSIC),
Glorieta de la Astronom{\'i}a s/n, 18008 Granada, Spain}

\author[0000-0002-6896-6085]{Jos\'e M. Torrelles}
\affiliation{Institut de Ci\`encies de l'Espai (ICE-CSIC), Can Magrans s/n, 08193, Bellaterra, Barcelona, Spain}
\affiliation{Institut d'Estudis Espacials de Catalunya (IEEC), Barcelona, Spain}

\author[0000-0002-3829-5591]{Josep M. Girart}
\affiliation{Institut de Ci\`encies de l'Espai (ICE-CSIC), Can Magrans s/n, 08193, Bellaterra, Barcelona, Spain}
\affiliation{Institut d'Estudis Espacials de Catalunya (IEEC), Barcelona, Spain}

\author[0000-0003-2775-442X]{Gabriele Surcis}
\affiliation{INAF - Osservatorio Astronomico di Cagliari, Via della Scienza 5, I-09047, Selargius, Italy}

\author[0000-0002-4987-5540]{Jeong-Sook Kim}
\affiliation{National Astronomical Observatories, Chinese Academy of Sciences, 20A Datun Road, Chaoyang District, Beijing, China}
\affiliation{Korea Astronomy and Space Science Institute, 776 Daedeokdaero, Yuseong, Daejeon 305-348, Republic of Korea}

\author[0000-0003-3863-7114]{Jorge Cant\'o}
\affiliation{Instituto de Astronom\'{i}a, Universidad Nacional Aut\'onoma de M\'{e}xico (UNAM), Apdo Postal 70-264, M\'{e}xico, D.F., Mexico}

\author[0000-0002-7506-5429]{Guillem Anglada}
\affiliation{Instituto de Astrof{\'i}sica de Andaluc{\'i}a (IAA-CSIC),
Glorieta de la Astronom{\'i}a s/n, 18008 Granada, Spain}

\author[0000-0003-4576-0436]{Salvador Curiel}
\affiliation{Instituto de Astronom\'{i}a, Universidad Nacional Aut\'onoma de M\'{e}xico (UNAM), Apdo Postal 70-264, M\'{e}xico, D.F., Mexico}

\author[0000-0002-2700-9916]{Wouter H.T. Vlemmings}
\affiliation{Department of Space, Earth and Environment, Chalmers University of Technology, Onsala Space Observatory, SE-439 92, Sweden}

\author[0000-0003-2862-5363]{Carlos Carrasco-Gonz\'alez}
\affiliation{Instituto de Radioastronom\'ia y Astrof\'isica (IRyA-UNAM),
Morelia, Mexico}

\author[0000-0002-4731-4934]{Adriana R. Rodr\'iguez-Kamenetzky}
\affiliation{Instituto de Astronom\'{\i}a Te\'orica y Experimental (IATE, CONICET-UNC), C\'ordoba, Argentina}

\author[0000-0002-0978-4775]{Soon-Wook Kim}
\affiliation{Korea Astronomy and Space Science Institute, 776 Daedeokdaero, Yuseong, Daejeon 305-348, Republic of Korea}

\author[0000-0002-2542-7743]{Ciriaco Goddi}
\affiliation{Instituto de Astronomia, Geof\'{\i}sica e Ci\^encias Atmosf\'ericas, Universidade de S\~ao Paulo, S\~ao Paulo, SP 05508-090, Brazil}
\affiliation{Dipartimento di Fisica, Universit\'a degli Studi di Cagliari, SP Monserrato-Sestu km 0.7, I-09042 Monserrato (CA), Italy}
\affiliation{INAF - Osservatorio Astronomico di Cagliari, Via della Scienza 5, I-09047, Selargius, Italy}
\affiliation{NFN, sezione di Cagliari, I-09042 Monserrato (CA), Italy}

\author[0000-0002-0230-5946]{Huib J. van Langevelde}
\affiliation{Joint Institute for VLBI ERIC, Oude Hoogeveensedijk 4, 7991 PD Dwingeloo, The Netherlands} \affiliation{Sterrewacht Leiden, Leiden University, Postbus 9513, 2300 RA Leiden, The Netherlands}

\author[0000-0002-3078-9482]{\'Alvaro Sanchez-Monge}
\affiliation{Institut de Ci\`encies de l'Espai (ICE-CSIC), Can Magrans s/n, 08193, Bellaterra, Barcelona, Spain}
\affiliation{Institut d'Estudis Espacials de Catalunya (IEEC), Barcelona, Spain}





\begin{abstract}

We have carried out ALMA observations of the massive star-forming region W75N(B), which contains the massive protostars VLA1, VLA2, and VLA3. Particularly, VLA2 is an enigmatic protostar
associated with a wind-driven H$_2$O maser shell, which has evolved from an almost isotropic outflow to a collimated one in just 20 years. The shell expansion seemed to be halted by an obstacle located to the northeast of VLA2. Here we present our findings from observing the 1.3~mm continuum and H$_2$CO and SiO emission lines. Within a region of $\sim$30$''$ ($\sim$39,000 au) diameter, we have detected 40 compact mm-continuum sources, three of them coinciding with VLA1, VLA2, and VLA3. While the H$_2$CO emission is mainly distributed in a fragmented structure around the three massive protostars, but without any of the main H$_2$CO clumps spatially coinciding with them,
the SiO is highly concentrated on VLA2, indicating the presence of very strong shocks generated near this protostar. The SiO emission is clearly resolved into an elongated structure ($\sim 0.6''\times0.3''$; $\sim 780$~au$\times$390~au) perpendicular to the major axis of the wind-driven maser shell. 
The structure and kinematics of the SiO emission are consistent with a {toroid and a wide-angle outflow surrounding a central mass of $\sim$10~M$_{\odot}$, thus supporting previous theoretical predictions regarding the evolution of the outflow}. 
Additionally, we have identified the expected location and estimated the gas density of the obstacle that is hindering the expansion of the maser shell.
\end{abstract}



\section{Introduction} \label{sec:intro}

Understanding how massive stars form and evolve remains a major challenge in astrophysics. 
One of the observational approaches to advance in this field has been to search for systems with disks and jets in massive protostars deeply embedded in their parental molecular clouds, trying to find similarities with the formation and evolution of low-mass stars.
Thus, several disk-jet systems in massive protostars are known today, indicating that they likely form in a similar way to low-mass stars, following the so-called monolithic core-accretion model, e.g., Cep A HW2  \citep[][]{Patel05,Jimenez07,Sanna17}, HH 80-81 \citep[][]{Girart18,Anez20}.
However, in the last years, several observational studies have indicated that during the early stages of evolution of massive protostars there may be episodic events of short duration (tens of years) associated with very poorly collimated or near-isotropic outflows \citep[e.g.,][]{Torrelles01,Kim13,Surcis14,Carrasco15,Bartk20}.
These results are surprising since, according to the core-accretion model, collimated outflows are already expected in the earliest phases of star formation, rather than outflows that expand without any preferential direction.  This expectation has been also confirmed with high-resolution observations of high-mass protostars in clustered  star forming regions \citep[e.g.,][]{Goddi20}. Therefore, it is still an open question whether the collimation and mass-ejection mechanisms in the early life of massive protostars differ from the case of low-mass protostars \citep[see, e.g.,][]{Carrasco21}.  This has important consequences for having a global vision of star formation, since mass ejection is related to the accretion process and then to the evolution of the protostar to its final mass. 

A very enigmatic case is the massive protostar W75N(B)-VLA2. Through a research program consisting of monitoring
for more than 20 years
an expanding water-vapor maser shell and the radio continuum emission associated with this source, it has been possible to observe the transformation of a short-lived, originally poorly collimated outflow from a massive protostar into a collimated jet,
as reported by 
\cite{Carrasco15}. This sudden change of the outflow morphology has been modeled by these authors in terms of an episodic, short-lived, originally isotropic, ionized wind that drives the observed maser shell around VLA2 and whose morphology evolves as a consequence of the interaction with a toroidal density stratification. However, these expanding motions are not symmetric because they seem to be halted by a possible dense obstacle at a distance of $\sim$0.2$''$ to the northeast of VLA2, as recently suggested by \cite{Surcis23} through a monitoring of the H$_2$O maser emission with the European VLBI Network (EVN).  A detailed description of the main properties of VLA2, as well as those of the other massive protostars in the high-mass star-forming region W75N(B), which is located at a distance of 1.3~kpc \citep[][]{Rygl12}, can be found in \cite{Surcis23}.

In this work, we present Atacama Large Millimeter/submillimeter Array (ALMA) continuum and H$_2$CO and SiO line observations at 1.3 mm wavelength toward the brightest mm core of the W75N(B) region, MM1 \citep[][]{Shepherd01,Minh10,R-K20,vanderWalt21}, which contains the massive protostars VLA1, VLA2, and VLA3 seen at radio continuum wavelengths \citep[][]{Hunter94,Torrelles97,R-K20}. These protostars have individual masses equivalent to B1-B0.5 spectral type stars \citep[e.g.,][]{Shepherd01}. The main goal of our ALMA observations was to test the presence of the molecular toroid around VLA2 predicted by previous works \citep[][]{Carrasco15}, as well as to study the distribution of gas and dust in the environment of these massive protostars. In a forthcoming paper we will analyze the emission from the other multiple molecular lines detected in this very active high-mass star-forming region.

\section{Observations} \label{sec:obs}

The ALMA observations in Band 6 ($\lambda$ = 1.1-1.4~mm; project 2019.1.00059.S; PI: Jeong-Sook Kim) were performed on four different sessions, and using two different array configurations: configuration C43-5 on 2021 April 29 (baselines 14--1261m) and 2022 August 2 (baselines 15--1301m), and configuration C43-8 on 2021 August 18 (baselines 92--8282m) and 2021 October 28 (baselines 63--8282m). The average observing time per session was $\sim$45 min, including calibration. The source J0006-0623 was used as flux and bandpass calibrator, while J2007+4029 was used as complex gain calibrator. The pointing position for the target W75N(B)-VLA2 was $\alpha$(J2000) = 20$^h$38$^m$36.486$^s$, $\delta$(J2000) = 42$^{\circ}$37$'$34.09$''$. The spectral setup consisted in six spectral windows with different bandwidths (BW) and velocity resolutions ($\Delta$V$_{ch}$), five of them centered on the rest frequencies (RF) of the molecular lines H$_2$CO [3(0,3)-2(0,2)], SiO [v=0, 5-4], C$^{18}$O [2-1], $^{13}$CO [2-1], $^{12}$CO [2-1], plus a continuum bandwidth (see Table~\ref{tab:setup}). The ALMA data were pipeline calibrated, reduced, and analyzed with CASA version 6.2.1.7 \citep[][]{casa22}. The continuum data (obtained from line free-channels of the spectral windows plus the continuum spectral window) were self-calibrated and their solutions applied also to the spectral line data. For imaging, we used task tclean  of CASA, with a Brigg's weighting of visibilities, and a robust parameter of 0.5 (unless otherwise noted), resulting in a synthesized beam of $\sim 0.17''\times 0.08''$ (PA $\simeq -2^{\circ}$). A high dynamic range for the continuum image of $\sim$1,100 (rms = 0.08~mJy~beam$^{-1}$) was reached after self-calibration (Fig.~\ref{fig:continuum}). Images presented in this paper were corrected by the primary beam response, which has a full width at half maximum of $\simeq$ 25$''$ at 230~GHz.

\begin{deluxetable*}{cccc}[htb]

\tablecaption{Spectral setup\label{tab:setup}}
\tablehead{\colhead{Spectral Window} & \colhead{RF} & \colhead{BW} & \colhead{$\Delta$V$_{ch}$} \\
\colhead{} & \colhead{(GHz)} & \colhead{(MHz)} & \colhead{(km~s$^{-1}$)}
}
\startdata
H$_2$CO [3(0,3)-2(0,2)]   & 218.222192 & 468.75 & 0.775 \\
SiO [v=0, 5-4]                    & 217.104980 & 468.75 & 0.779 \\
C$^{18}$O [2-1]                & 219.560358 & 234.38 & 0.385 \\
$^{13}$CO [2-1]                & 220.398684 & 234.38 & 0.384 \\
$^{12}$CO [2-1]                & 230.538000 & 468.75 & 0.734 \\
Continuum       & 231.805970 & 1,875   & 40.443 \\
\enddata
\end{deluxetable*}

\section{Results} \label{sec:res}

\subsection{Dust continuum emission} \label{sec:dust}

\begin{figure}[ht]
\includegraphics[width=1\textwidth]{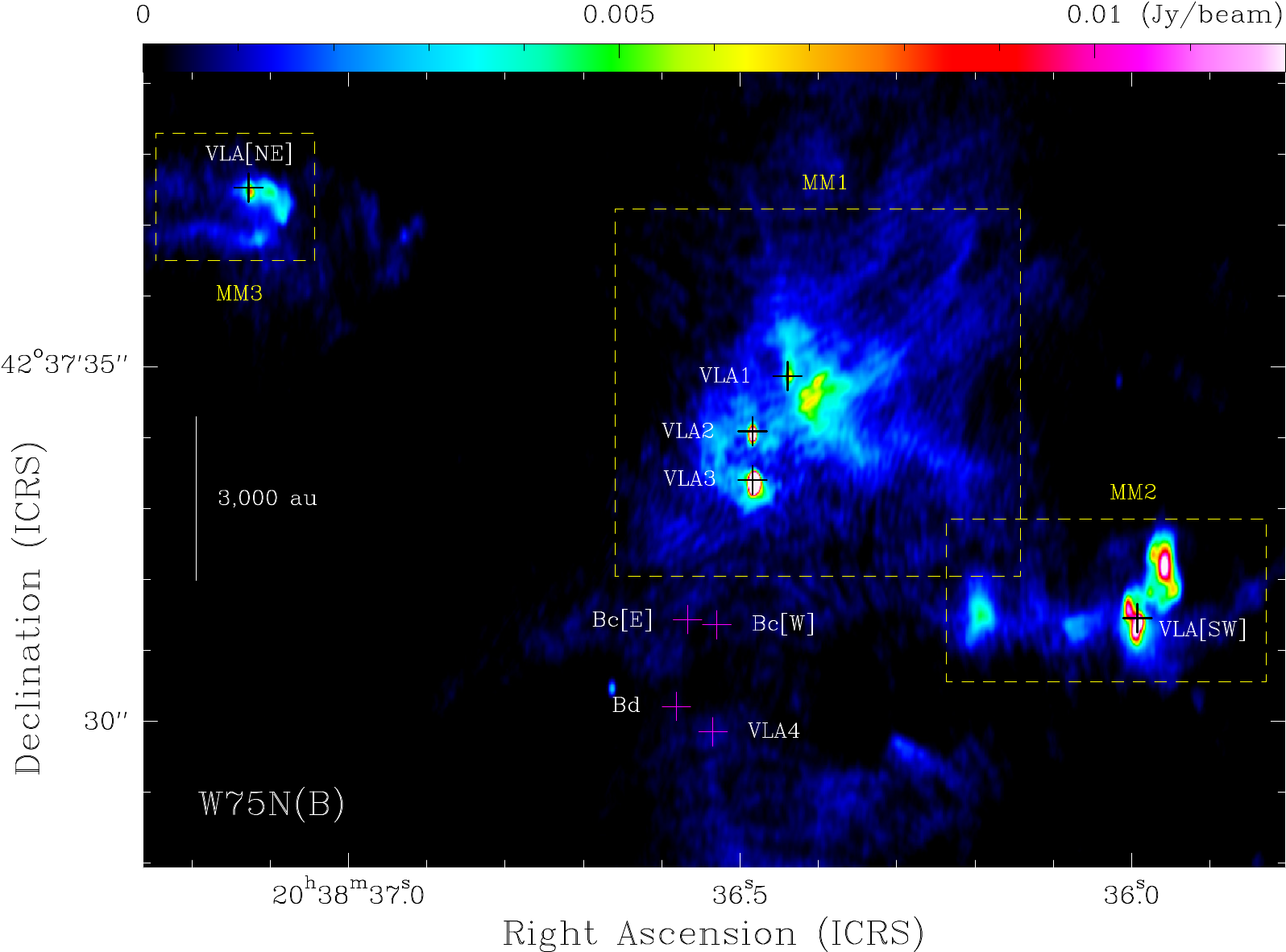}
\caption{ALMA 1.3 mm continuum emission image of the high-mass star-forming region W75N(B). The intensity range of the image has been saturated at 15 mJy~beam$^{-1}$ 
(beam $= 0.17''\times0.08''$, PA = $-2^{\circ}$; rms $\simeq 0.08$~mJy~beam$^{-1}$) 
to distinguish the faintest continuum sources, given the high dynamic range of the image ($\sim$1,100). The maximum  of the image (coincident with VLA3) is 97~mJy~beam$^{-1}$.
The position of the radio continuum sources detected in the region at cm wavelengths, including the massive protostars VLA1, VLA2, and VLA3, are labeled and indicated by crosses, while dotted rectangles indicate the regions MM1, MM2, and MM3 where continuum emission was previously reported at 1.3~mm with ALMA, but with a beam of $1.73''\times0.86''$,  PA = $-4^{\circ}$ \citep{R-K20}. Table~\ref{tab:possources} lists the positions of all the mm continuum sources detected in the region (see \S\ref{sec:dust}, \S\ref{ap:sources}, and Fig.~\ref{fig:sources}). 
\label{fig:continuum}}
\end{figure}

Figure~\ref{fig:continuum} shows the continuum image obtained towards the central part of the W75N(B) region. The millimeter cores MM1, MM2, and MM3 \citep[previously observed with a beam of $\simeq$ 1$''$--2$''$:][]{Minh10,R-K20,vanderWalt21}, are now resolved into multiple compact millimeter sources. All known radio continuum sources at centimeter wavelengths in this region \citep[labeled and indicated by crosses in Fig.~\ref{fig:continuum}; from][]{R-K20}, except Bc[E], Bc[W], Bd, and VLA4, have associated continuum emission at millimeter wavelengths, further reinforcing that VLA1, VLA2, VLA3, VLA[SW], and VLA[NE] are embedded protostars. The full width at half maximum (FWHM) of these continuum sources is unresolved by our beam.
On the other hand, the fact that Bc[E], Bc[W], Bd, and VLA4 have no continuum emission at millimeter wavelengths (Fig.~\ref{fig:continuum}) gives support to the interpretation by \cite{Carrasco10} and  \cite{R-K20}, through proper motion measurements, that these four objects are not embedded protostars, but obscured Herbig-Haro (HH) objects excited by the massive protostar VLA3. 

In order to better distinguish weak compact mm continuum sources from extended emission present in the region (see Fig.~\ref{fig:continuum}), we obtained images that exclude baselines shorter than 200~k$\lambda$, thus filtering out large-scale structures, in a similar way as in \cite{Girart18} and \cite{Busquet19}.  
With this baseline restriction, 40 unresolved continuum sources have been identified in a region of $\sim$30$''$ ($\sim$39,000 au) diameter centered on VLA2,  and with flux densities in the range of $\sim$0.4--90~mJy~beam$^{-1}$ (Table~\ref{tab:possources},  Fig.~\ref{fig:sources}).  We are proposing, in particular, that one of these detected sources (ID 32 in Table~\ref{tab:possources}) is the powering source of the highly collimated bipolar jet that we have detected $\sim$6$''$ ($\sim$7,800 au) northeast of VLA2 (see Appendix~\ref{ap:siojet} and Fig.~\ref{fig:siojet}), and that will be {discussed} in detail in a forthcoming paper.

\subsection{H$_2$CO line emission} \label{sec:h2co}

\begin{figure}[ht]
\includegraphics[width=1\textwidth]{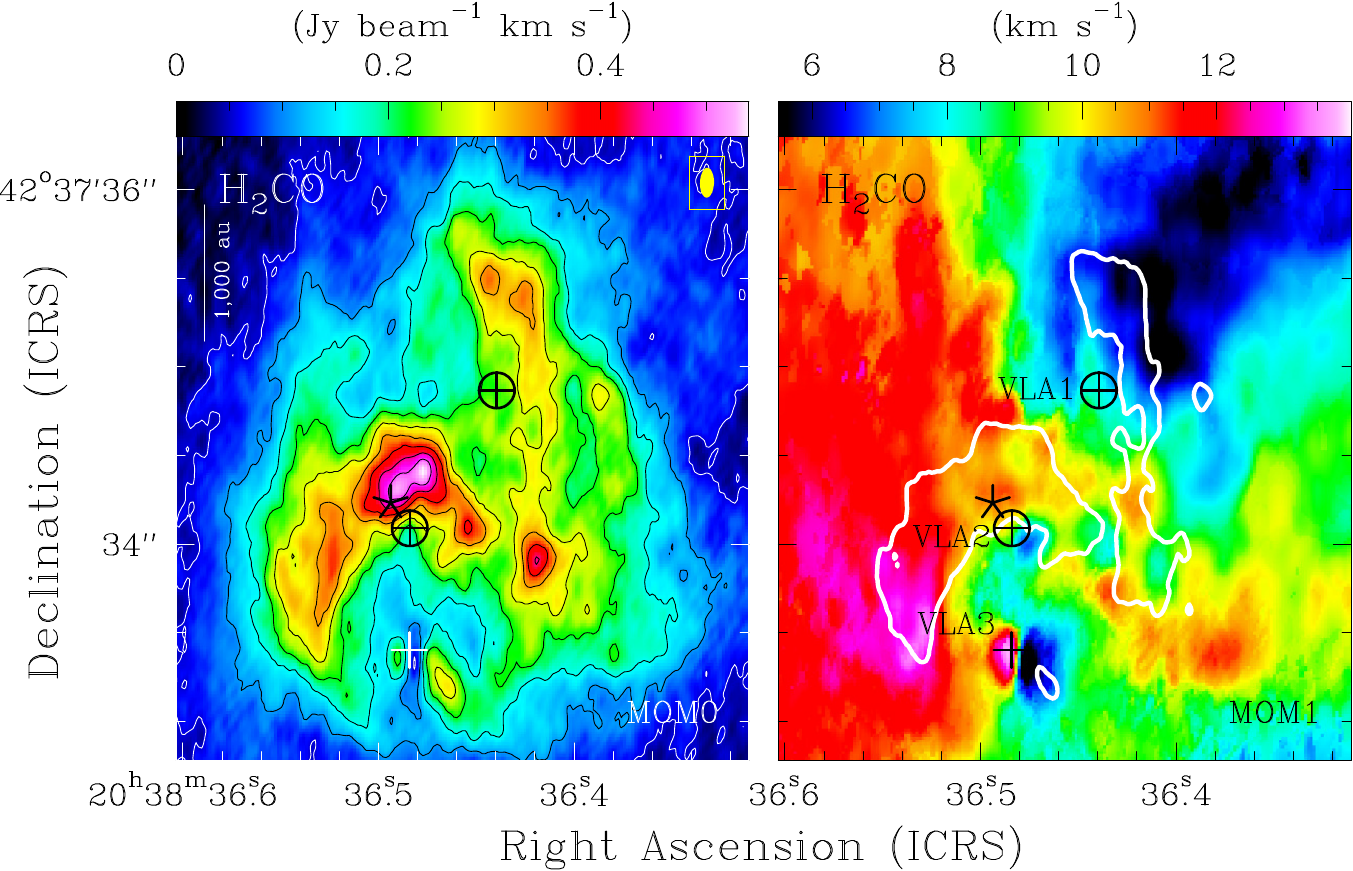}
\caption{{\it Left panel (a):} Integrated intensity image (moment of order 0) of the 
H$_2$CO [3(0,3)-2(0,2)] emission line in the velocity range $V_{\rm LSR} \simeq -4$ to +23~km~s$^{-1}$.
Contour levels are 10, 20, 30, 40, 50, 60, 70, 80, 90 per cent of the maximum integrated intensity of the image, 0.54~Jy~beam$^{-1}$~km~s$^{-1}$. The synthesized beam is plotted as yellow ellipse at the upper right corner of the image.
{\it Right panel (b):} velocity field image (moment of order 1). To obtain this moment-1 image, we applied a 4$\sigma$ threshold (4~mJy~beam$^{-1}$) to the individual spectral channels. { The white line traces the half-power contour of the MOM0 image in left panel.
Ringed cross symbols for VLA1 and VLA2 indicate that these are the activity centers of strong H$_2$O maser emission in the W75N(B) star-forming region \citep[see, e.g.,][]{Surcis23}.
The five-pointed star in both panels marks the location of the northeastern masers components in the VLA2 H$_2$O maser shell, where the expansion appears to have halted \citep{Surcis23}}.
\label{fig:h2co}}
\end{figure}

The emission of the formaldehyde molecule is known to be a good tracer of dense gas ($n[{\rm H}_2] \gtrsim 5 \times 10^5 -10^6$~cm$^{-3}$) over a wide range of temperatures ($T_K \simeq 10-300$~K) in the cores of molecular clouds \citep[e.g.,][]{Mangum93,Shirley15}. The integrated emission image obtained from the H$_2$CO [3(0,3)-2(0,2)] line in the velocity range $V_{\rm LSR} \simeq  -4$ to +23~km s$^{-1}$ (Fig.~\ref{fig:h2co}a) shows a clumpy structure over a plateau of relatively weak emission covering a region of $\sim 3-4''$ ($\sim 3,900-5,200$ au). None of the main H$_2$CO clumps coincide with the position of any of the massive protostars VLA1, VLA2, or VLA3. 
The velocity field  of the H$_2$CO emission (Fig.~\ref{fig:h2co}b) shows that the { greatest velocity differences} are found { along} the northwest-southeast direction at angular scales of $\sim$3$''$ ($\sim$3,900 au), with blueshifted ($V_{\rm LSR}$ $\simeq$ +6~km s$^{-1}$)  and redshifted velocities ($V_{\rm LSR}$ $\simeq$ +13~km s$^{-1}$) to the northwest and  southeast of VLA2, respectively \citep[ambient velocity $V_{\rm LSR}$ $\simeq$ +10~km s$^{-1}$, e.g.,][]{Surcis23}. { This direction also coincides with the orientation of the strongest H$_2$CO emission (Fig.~\ref{fig:h2co}b) and} consistent with the velocity field previously observed with the Submillimeter Array (SMA) and ALMA through different molecular species (e.g., H$_2$CO, CS, SO$_2$, CH$_3$OH), at scales of $\sim 4''$ and angular resolutions of  $\sim 1''-2''$ \citep[][]{Minh10,R-K20,vanderWalt21}. In particular, \cite{vanderWalt21} suggested that the observed velocity shift
is due to infalling gas onto a disk-like structure surrounding one of the massive protostars, but without being able to specify which one, probably due to an insufficient  angular resolution in their observations ($\sim$1$''$). More recently, \cite{Zeng23} conclude that the gas infall proceeds along the magnetic field lines,
as the gravitational force dominates over the magnetic field.

We suggest that the distribution of the clumpy H$_2$CO  structure near but around VLA1, VLA2, and VLA3 could be due either to residual dense molecular gas fragments after the  formation of these massive protostars, or to material still infalling toward the center of the region. 
This is supported by the fact that none of the main H$_2$CO clumps observed around the massive protostars have an associated mm continuum source (see Figs.~\ref{fig:sources}, \ref{fig:h2co}a, Table~\ref{tab:possources}).
The clumps observed in H$_2$CO have a relatively low mass of gas. For example, for the most intense clump (located $\sim0.2''-0.3''$ northeast of VLA2),  assuming a density 
$n({\rm H}_2) \gtrsim (5\times 10^5)-10^6$~cm$^{-3}$, and adopting a diameter size of 0.5$''$ (650~au) from the emission image (Fig.~\ref{fig:h2co}a), we roughly estimate a molecular mass $M({\rm H}_2) \gtrsim 4-8\times10^{-4}$~M$_{\odot}$. In addition, 
given its proximity to 
VLA2, we consider it plausible that this particular dense gas could be related to the obstacle predicted by \cite{Surcis23} to halt the expanding motions of the H$_2$O masers observed to the northeast of VLA2 (see Fig.~\ref{fig:h2co} and discussion in \S\ref{sec:halted}). 

We also identify an H$_2$CO disk-like structure centered on VLA3 with a velocity gradient in the east-west direction (see Fig.~\ref{fig:h2co}), therefore perpendicular to the VLA3 radio jet oriented in the north-south direction \citep[][]{Carrasco10,R-K20}.  In addition, the H$_2$CO line is seen in absorption toward VLA3 (absorption is also observed in SiO at this location; see Fig.~\ref{fig:absorption}). These results on VLA3 will be presented and discussed in a forthcoming paper.

\subsection{General distribution of the SiO line emission} \label{sec:sio}

\begin{figure}[ht]
\includegraphics[width=1\textwidth]{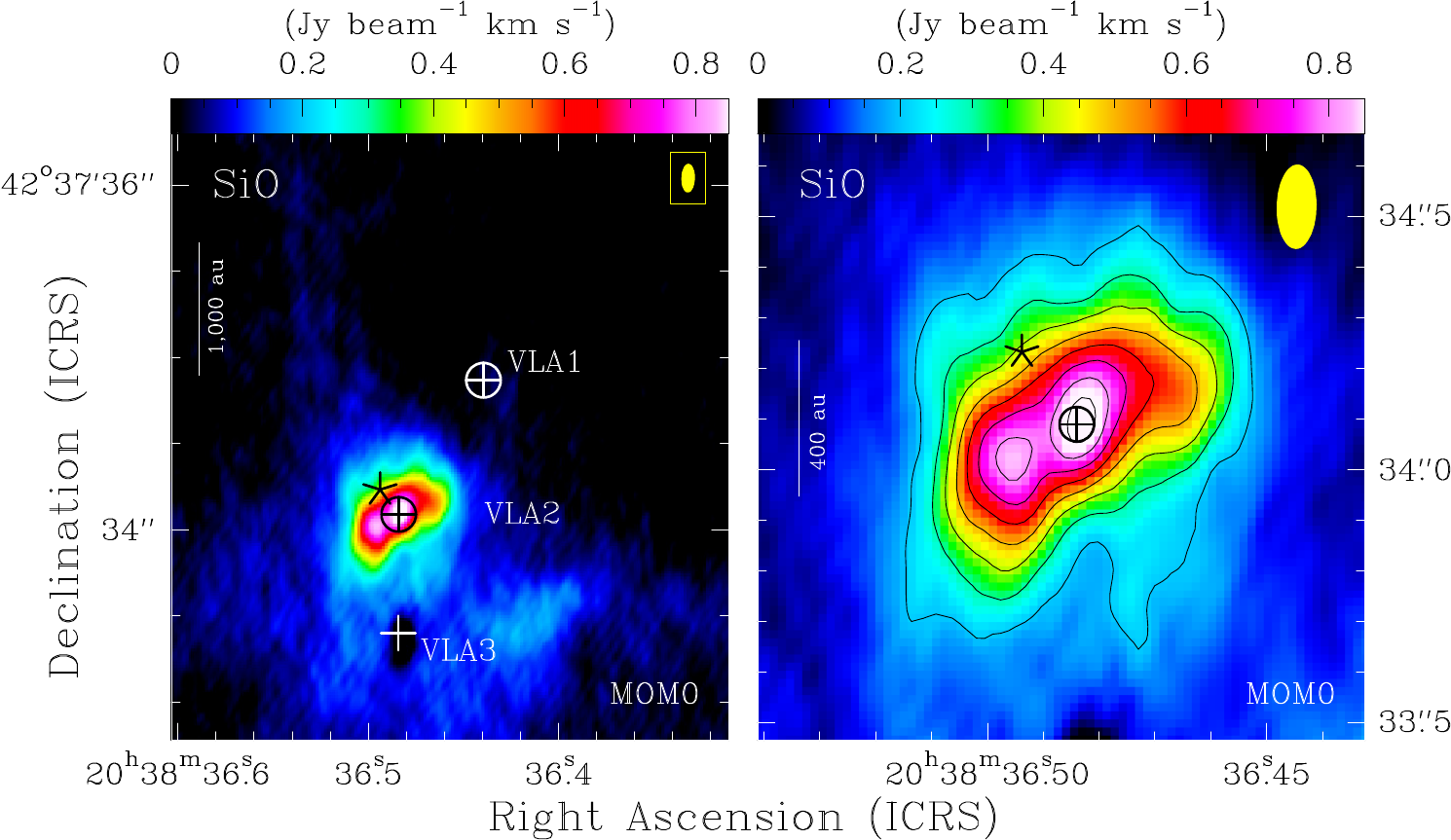}
\caption{{\it Left panel (a):} Integrated intensity image (moment of order 0) of the 
SiO [v=0, 5-4] emission line in the velocity range V$_{\rm LSR}$ $\simeq$ --20 to +28 km s$^{-1}$.
The displayed field of view is the same as the one shown in Fig.~\ref{fig:h2co}a for H$_2$CO. {\it Right panel (b):} zoom-in showing the integrated SiO emission around VLA2. Contour levels are 0.2, 0.3, 0.4, 0.5, 0.6, 0.7, 0.8, and 0.9~Jy~beam$^{-1}$~km~s$^{-1}$. { All symbols used in this figure are the same as those shown in Fig.~\ref{fig:h2co}.}
\label{fig:sio}}
\end{figure}

The spatial and kinematic distribution of silicon monoxide emission is completely different from that of formaldehyde discussed in the previous section (\S\ref{sec:h2co}). In Figure~\ref{fig:sio}a  we show the image of the integrated intensity of the SiO [v=0, 5-4] emission line covering the same spatial region as that of H$_2$CO in Figure~\ref{fig:h2co}a. From the comparison of these two images (Figs.~\ref{fig:h2co}a and \ref{fig:sio}a), we see that, while the H$_2$CO emission is distributed in several bright molecular clumps close, but not coinciding with any of the three massive protostars (VLA1, VLA2, VLA3), the SiO emission is strongly and almost exclusively concentrated around VLA2. 
This remarkable behavior related to the strong concentration of SiO  emission on VLA2 had already been reported by \cite{Minh10} through SMA observations, but without resolving its spatial distribution. With our ALMA observations we now resolve the SiO emission around VLA2 into a northwest-southeast elongated structure with a size of $\sim0.6''\times0.3''$ ($\sim$780~au$\times$390~au; PA $\simeq$ $-$53$^{\circ}$), as seen in Figure~\ref{fig:sio}b. The range of velocities with SiO emission observed within this elongated structure surrounding VLA2 is $V_{\rm LSR} \simeq -20$ to +28~km s$^{-1}$, which is significantly larger than that of H$_2$CO emission throughout all the region containing the three massive protostars (\S\ref{sec:h2co}).

The SiO emission is a well-known tracer of shocked gas, since the molecule is present in the gas phase in regions undergoing shocks with velocities of $\ga 25$ km s$^{-1}$ \citep[e.g.,][]{Schilke97}. Moreover, the SiO lines are high-density tracers and, for instance, the critical density of excitation for the SiO(5-4) transition is $\simeq 1.5\times10^{6}$~cm$^{-3}$ \citep{Huang22}. 
These extreme properties of the SiO excitation, together with the distribution of its emission described above, lead us to conclude the presence of very strong shocks around VLA2. We think that this is a consequence of the high outflow activity that VLA2 has, with very intense H$_2$O masers expanding at several tens of km s$^{-1}$ driven by strong repetitive, short-lived ionized winds from the central source \citep[][]{Torrelles03,Kim13,Carrasco15,Surcis23}. All these properties observed in VLA2 are what make this massive protostar unique in comparison with the other two massive protostars in the region (VLA1 and VLA3). 
In \S\ref{sec:toroid} we discuss and interpret the SiO structure with its kinematics in terms of a molecular toroid surrounding VLA2, together with a wide-angle outflow.
\section{Discussion}

\subsection{The SiO toroid surrounding W75N(B)-VLA2} \label{sec:toroid}

The elongated SiO structure ($\sim$780~au$\times$390~au, PA $\simeq$ $-$53$^{\circ}$; \S\ref{sec:sio}, Fig.~\ref{fig:sio}b)
that surrounds VLA2 is approximately perpendicular to the major axis of the expanding H$_2$O maser shell \citep[PA $\simeq$ 45$^{\circ}$;][]{Surcis14,Surcis23}, which, in its turn, is shock-excited and driven by the VLA2 ionized jet \citep[PA $\simeq$ 65$^{\circ}$;][]{Carrasco15}. We interpret this SiO molecular structure as the toroid whose existence was hypothesized and modeled by \cite{Carrasco15}, as the collimating agent of an episodic, originally isotropic ionized wind associated with VLA2. These authors explain the observed evolution of a non-collimated outflow to a collimated outflow in VLA2 as due to a toroidal environmental density stratification, seen almost edge-on, with densities $\sim$5$\times$10$^{7}$~cm$^{-3}$ at distances of $\sim$28~au from the central massive protostar. 
This seems to be consistent with the fact that SiO(5-4) emission, which traces densities $\gtrsim$ 1.5$\times$10$^{6}$~cm$^{-3}$, is observed up to distances of $\sim$390~au from VLA2.
Assuming this critical density of SiO as the minimum density for the entire toroid-like structure and given its observed size, a lower limit for the molecular mass  $M({\rm H}_2) \gtrsim 2\times 10^{-3}$~M$_{\odot}$ is estimated. An independent estimate for a lower limit of the total mass (gas+dust) in this region can be obtained from the observed dust continuum emission of the unresolved VLA2 source (see \S\ref{ap:sources}), under the assumption of optically thin emission and using Eq.~\ref{eq:massdust} (see \S\ref{ap:sources}). Adopting $T_d$ = 300~K (a reasonable value for gas temperatures near the massive object,  considering the presence of strong H$_2$O maser emission nearby, which implies that the gas should be at high-temperatures), we obtain a total mass (gas+dust) $\gtrsim 5\times 10^{-2}$~M$_{\odot}$ ($\gtrsim 51~M_J$; Table~\ref{tab:possources}). 

\begin{figure}[ht]
\includegraphics[width=1\textwidth]{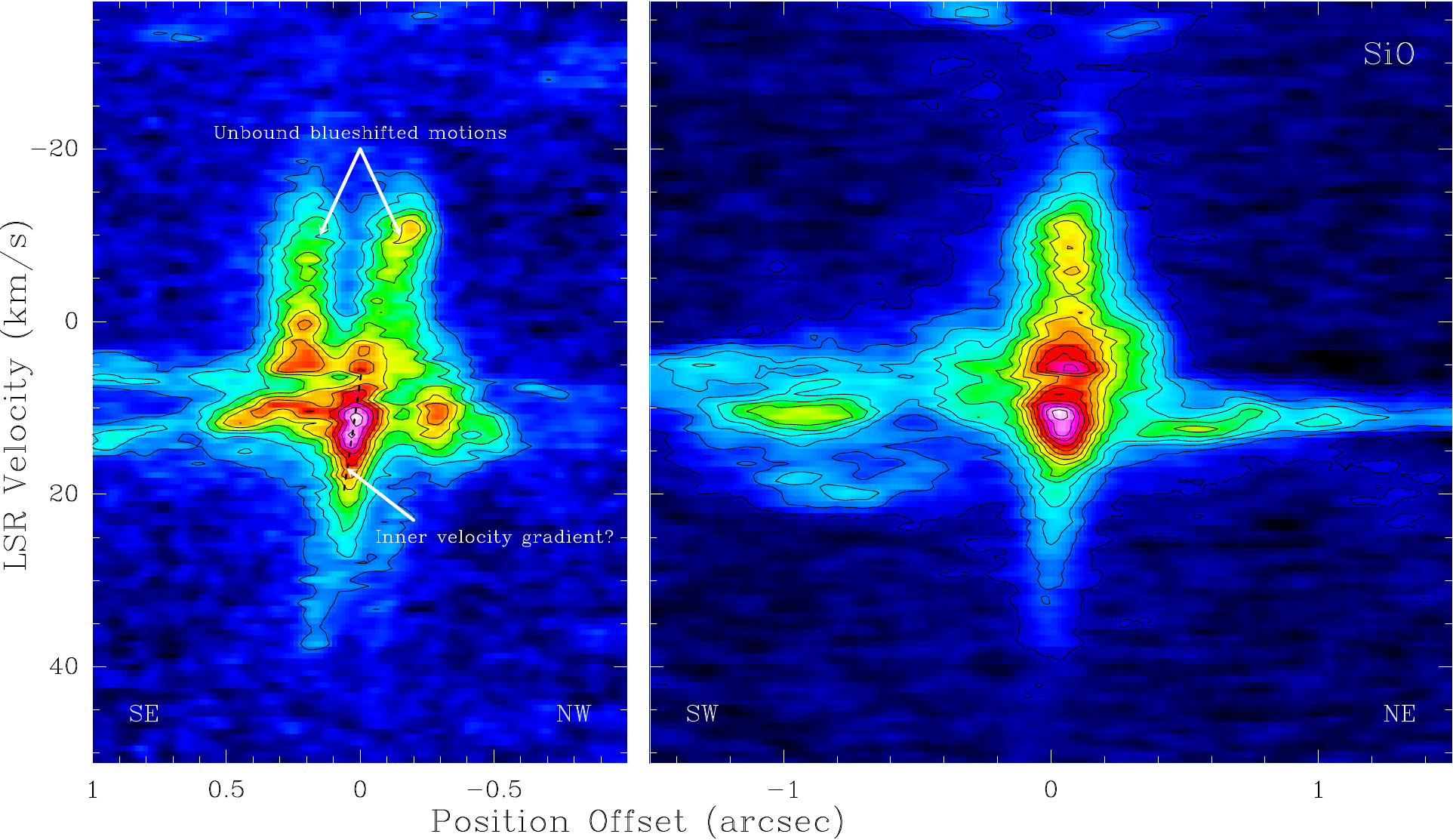}
\begin{center}
\includegraphics[width=0.5\textwidth]{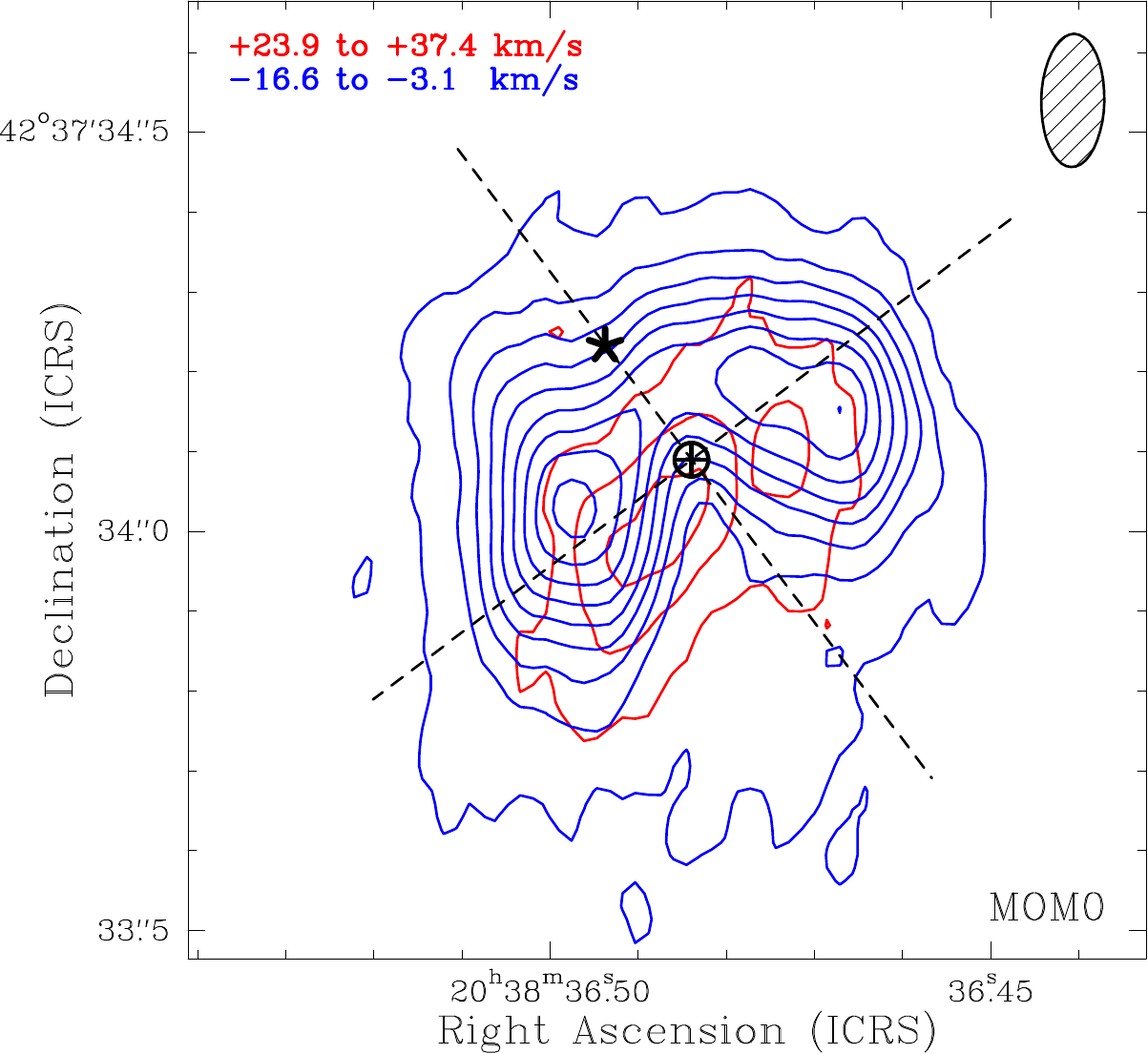}
\end{center}
\caption{{ Diagrams and image showing the various velocity components of the VLA2 SiO toroid.} {\it Upper left panel (a):} Position-velocity diagram of the SiO emission along the major axis of the elongated structure (PA $\simeq$ $-$53$^{\circ}$; { dashed line in Fig.~\ref{fig:siokine}c}) with an averaging width of 0.14~arcsec. 
The first contour level and the increment step are $3\sigma$, where $\sigma=1.4$ mJy~beam$^{-1}$ (the rms of the diagram).
The maximum of the diagram is 44~mJy~beam$^{-1}$. The two `rabbit-ear-shaped" structures tracing unbound blueshifted motions are indicated. These structures originate from two peaks seen in the high-velocity blueshifted lobe (see \S\ref{sec:toroid} and Fig.~\ref{fig:siokine}c). The dotted line indicates a possible inner velocity gradient in the SiO toroid. {\it Upper right panel (b):} same as upper left panel, but for the minor axis (PA $\simeq$ $-$143$^{\circ}$; { dashed line in Fig.~\ref{fig:siokine}c}) with an averaging width of 0.6~arcsec. 
The first contour level is $3\sigma$ and the increment step is $6\sigma$, where $\sigma=0.38$ mJybeam$^{-1}$. The maximum of the diagram is 32~mJy~beam$^{-1}$.
Offset positions are with respect to the maximum SiO integrated intensity position (Fig.~\ref{fig:sio}b).
{\it Lower panel (c):} Integrated intensity image (moment of order 0) of the blueshifted and redshifted SiO in the high velocities ranges $V_{\rm LSR} = -16.6$ to $-3.1$~km~s$^{-1}$ and $V_{\rm LSR}$ = +23.9 to +37.4~km~s$^{-1}$, respectively. 
The first contour level is $4\sigma$ and the increment step is $6\sigma$, where $\sigma=5$~mJy~beam$^{-1}$~km~s$^{-1}$. { Symbols are as in Fig.~\ref{fig:h2co}.}
\label{fig:siokine}}
\end{figure}

Disk-like structures have been also reported in SiO emission surrounding other massive protostars \citep[e.g.,][]{Gomez99,Matthews10,Maud18}. In particular, \cite{Maud18}, through ALMA 1.3 mm observations, report a rotating SiO disk and a disk wind from the massive protostar G17.64+0.16 (probably an O-type star). 
The radius of the SiO disk structure in that object extends up to $\sim$600~au, similar
to that of the SiO toroid  observed around VLA2 (in terms of nomenclature we refer as toroids to disk-like structures with a size of $\sim$1,000 au). The very large velocity components observed in SiO within the VLA2 toroid ($V_{\rm LSR}\simeq -20$ to +28 km s$^{-1}$)  are consistent with the velocities of the shocks needed to produce SiO as proposed by  \cite{Maud18} for the disk (toroid) of G17.64+0.16. {Similarly to G17.64+0.16, the shocks in VLA2 may originate at the inner edge and surface interface of the toroid by the fast expansion of the ionized wind and outflow \citep[$\gtrsim$ 30~km~s$^{-1}$,][]{Surcis14,Carrasco15}.}

Figure~\ref{fig:siokine}a shows the position velocity (PV) diagram of the SiO emission along the major axis of the toroid-like structure (PA = $-$53$^{\circ}$) with an averaging width of 0.14$''$ (selected to optimize the visualization of the SiO kinematics). A velocity gradient of $\sim$6~km s$^{-1}$ is observed in the inner and most intense part of the SiO structure at angular scales of $\sim$0.1$''$ ($\sim$130~au), with the more blueshifted velocities to the northwest and the more redshifted velocities to the southeast. This velocity shift is in the same sense as the velocity shift observed northwest-southeast in H$_2$CO at larger scales \citep[$\sim 4''$, Fig.~\ref{fig:h2co}b; see also][]{Minh10} and in other molecular species \citep[e.g., SO$_2$, CH$_3$OH;][]{R-K20}. If the observed velocity shift of 6 km s$^{-1}$ is due to rotating motions at the inner parts ($\sim$130~au) of the toroid, those motions could be gravitationally bound by a central mass of $\sim$10~M$_{\odot}$ {assuming Keplerian conditions}, consistent with the expected mass of a B1-B0.5 espectral type star for VLA2 \citep[][]{Shepherd01}. However, it seems clear that observations at submillimetre wavelengths, with higher angular resolution than currently available in this work, are needed to resolve in more detail the kinematic behavior of molecular gas in the internal parts of the toroid (at angular scales of $\lesssim 0.1''$).

The  weakest SiO emission displays high blueshifted velocities, reaching up to $V_{\rm LSR}$ $\simeq -17$~km s$^{-1}$, which appear as two rabbit-ear-shaped structures in the PV diagram along the major axis of the toroid (Fig.~\ref{fig:siokine}a), and located symmetrically on both sides ($\sim\pm 0.2''$) of VLA2.
 Additionally, there is high redshifted velocity emission, with velocities up to $V_{\rm LSR}$ $\simeq +37$~km$^{-1}$, which is more apparent in the PV diagram along the minor axis of the SiO structure with an averaging width of $0.6''$ (Fig.~\ref{fig:siokine}b).
To explain these motions, which amount to approximately 27~km s$^{-1}$ relative to the ambient velocity ($V_{\rm LSR}$ = +10~km s$^{-1}$; e.g., Surcis et al. 2023), and considering the distances of 0.2$''$ (260~au), a central mass of around 210~M$_{\odot}$ would be required. However, such a mass is not observed, leading us to conclude that these high velocities traced by the weakest SiO emission represent unbound motions of an outflow originating from the outer parts ($\sim\pm 0.2''$) of the toroid.

Figure~\ref{fig:siokine}c shows contour maps of the blueshifted ($V_{\rm LSR} = -16.6$ to $-3.1$ km s$^{-1}$) and redshifted ($V_{\rm LSR}$ = +23.9 to +37.4 km s$^{-1}$) SiO integrated emission, with the blueshifted lobe being stronger than the redshifted one. This figure allows us to identify the source of the two rabbit-ear-shaped structures observed in Figure~\ref{fig:siokine}a as originating from the two peaks seen in the blueshifted lobe. These two peaks are separated by $\sim$0.4$''$, with the VLA2 source located between them.

It is worth noting that the SiO blueshifted lobe is slightly displaced towards the northeast of VLA2, while the redshifted lobe is slightly displaced towards the southwest (Fig.~\ref{fig:siokine}c). This displacement is consistent with the direction observed in the CO bipolar outflow at parsec scales \citep[blueshifted to the northeast, redshifted to the southwest; e.g.,][]{Davis98a,Davis98b,Shepherd03}, indicating that the SiO outflow is being incorporated into the larger-scale molecular outflow, similar to the case of G17.64+0.16 \citep[][]{Maud18}. Furthermore, the morphological distribution of the two SiO outflow lobes (Fig.~\ref{fig:siokine}c), which spatially surround VLA2, indicates the presence of a wide-angle outflow associated with this massive protostar.

Finally, we have considered the possibility that the two rabbit-ear-shaped structures that can be distinguished in Figure~\ref{fig:siokine}a are due to the presence of two protostars embedded in the SiO toroid, one of which being VLA2. However, the fact that no other continuum source is detected, despite the high sensitivity of the data, together with the distribution of the two SiO outflow lobes around VLA2, makes this possibility very unlikely.

\subsection{A wind emerging from the toroid, interacting with a high-density gas obstacle} \label{sec:halted}

As metioned above, \cite{Carrasco15} showed that both, the VLA2 radio continuum source and the maser shell around it evolved in $\sim$20~years from a compact source into an elongated one in the northeast-southwest direction. They successfully modeled this behavior as due to the interaction of an episodic, short-lived isotropic ionized wind with a toroidal density stratification around the source. And we now find this toroidal structure, traced by the SiO emission.

Recently, \cite{Surcis23} found that the northeast part of the shell traced by the H$_2$O masers has stopped, and they suggest  that this is due to the interaction of the shell with a high-density gas obstacle. An analytical and numerical simulation of such a situation by \cite{Canto06} show that, indeed, the interaction of a moving shell with a very dense medium can slow down the shell and stop  its expansion completely. The detection of a strong H$_2$CO clump to the northeast of VLA2 as reported in the present paper (\S\ref{sec:h2co}, Fig.~\ref{fig:h2co}a) seems to be consistent with the suggestion of \cite{Surcis23}.

Following this idea, we can estimate the required gas density of the obstacle by assuming a balance between the thermal pressure of the gas in the clump and the ram pressure of the wind that drives the shell. That is,

\begin{equation} \label{eq:rampress}
nkT_k = \frac{\dot{M}_{w} V_{w}}{4\pi R^2},
 \end{equation}
where $n$ is the density of the obstacle, $T_k$ its kinetic temperature, $k$ the Boltzmann constant, $\dot{M}_{w}$ and $V_{w}$ the mass-loss rate and terminal velocity of the isotropic wind, respectively, and $R$ the distance between the observed halted H$_2$O masers and the powering source of the wind.
Solving for the density of the clump we find,
 
 \begin{equation} \label{eq:densityclump}
\left[\frac{n}{\rm cm^{-3}}\right] = 1.6\times10^9 {\Large \frac{\left[\frac{\dot{M}_{w}}{\rm 10^{-7} M_{\odot}{yr}^{-1}}\right] \times \left[\frac{V_{w}}{\rm 100~km~s^{-1}}\right]}{\left[\frac{T_{k}}{\rm 10~K}\right] \times \left[\frac{R}{\rm 100~au}\right]^2}}.
 \end{equation}
Assuming for $\dot{M}_{w}$ and $V_w$ the values quoted above from \cite{Carrasco15}, together with $T_k$ = 300~K and R = 260~au \citep[$\sim 0.2''$;][]{Surcis23}, we found $n \simeq 3.5\times 10^7-3.0\times10^8$~cm$^{-3}$.
This is a very high density, but consistent with the fact that this obstacle is located within a very high-density region, close to the observed H$_2$CO clump to the northeast of VLA2 (Fig.~\ref{fig:h2co}a) and to the outer parts of the SiO toroid (Fig.~\ref{fig:sio}b). In addition, we should also consider that the density we estimated for the obstacle to halt the expanding motions of the H$_2$O masers to the northeast of VLA2 is likely to be an upper limit to the actual density.
This is so because: 1) we have calculated the gas pressure (left expression of Eq.~\ref{eq:rampress}) from the kinetic temperature only, but if the clump has turbulent motions  \citep[which is most likely the case; see][]{Surcis23} the actual pressure will be higher;  
2) the ram pressure (right expression in Eq.~\ref{eq:rampress}) assumes that the shock is normal to the direction of the wind, but any deviation from this condition will produce a lower ram pressure;
3) to estimate the density in Eq.~\ref{eq:densityclump}, we have used the projected distance between the obstacle and the source, but the actual linear distance might be  longer. 

\section{Summary} \label{sec:sum}

Observations of the W75N(B) massive star-forming region using ALMA at a wavelength of 1.3 mm (beam $= 0.17''\times0.08''$, PA = $-2^{\circ}$) are presented, focusing on the continuum, H$_2$CO, and SiO lines. This region contains several massive protostars, including VLA1, VLA2, and VLA3, which are clustered within an area of $\sim$2$''$ in size ($\sim$2,600 au). The most relevant results are summarized as follows: 

\begin{itemize}

\item{We detected 40 compact 1.3 mm continuum sources within an area of $\sim 30''$ diameter ($\sim 39,000$ au), with three of them coinciding with VLA1, VLA2, and VLA3.  The majority of these sources are new detections and are likely indicative of low- or intermediate-mass YSOs.}

\item{The H$_2$CO emission is distributed in a clumpy structure around VLA1, VLA2, and VLA3. In contrast, the SiO emission is highly concentrated almost exclusively on VLA2, implying the presence of intense shocks in the vecinity of this particular massive protostar. The SiO emission outlines an elongated structure that is perpendicular to the major axis of the wind-driven shell of H$_2$O masers and its associated radio jet. We identify this structure as the toroidal structure that was previously theorized to explain the outflow's evolution from an almost isotropic outflow to a collimated one. }

\item{The observations provide a plausible scenario in which we are observing an SiO toroid and a wide-angle outflow surrounding VLA2. The wide-angle SiO outflow originates from the outer regions of the toroid, being incorporated into the large-scale bipolar molecular outflow. The gas located in the inner regions of the toroid might be rotating, although more precise observations are required to confirm this.}

\item{The location of the obstacle that was previously predicted to stop the expansion of the wind-driven H$_2$O maser shell has been identified. It is situated  between the outer parts of the SiO toroid and one of the main H$_2$CO clumps detected nearby, toward the northeast of VLA2. Additionally, this determination involves estimating the necessary gas density  to impede the shell's expansion toward the northeast.}

\end{itemize}

\subsection*{acknowledgments}
This paper makes use of ALMA data ADS/JAO.ALMA\#2019.1.00059.S. ALMA is a partnership of ESO (representing its member states), NSF (USA) and NINS (Japan), together with NRC (Canada), MOST and ASIAA (Taiwan), and KASI (Republic of Korea), in cooperation with the Republic of Chile. The Joint ALMA Observatory is operated by ESO, AUI/NRAO and NAOJ.
The data processing for this paper has been carried out using the Spanish Prototype of an SRC (SPSRC) service and support, funded by the Spanish Ministry of Science, Innovation and Universities, by the Regional Government of Andalusia, by the European Regional Development Funds and by the European Union NextGenerationEU/PRTR. 
{J.C. acknowledges the support of DGAPA/UNAM (Mexico) project IG100422. C.C.-G. acknowledges support from UNAM DGAPA-PAPIIT grant IG101321 and from CONACyT Ciencia de Frontera project ID 86372}. G.A and J.F.G. acknowledge support from grants PID2020-114461GB-I00 and CEX2021-001131-S, funded by MCIN/AEI/10.13039/5011000110. 
J.M.G, A.S-M., and J.M.T. acknowledge partial support from the PID2020-117710GB-I00 grant funded by MCIN/AEI/10.13039/501100011033, and by the program Unidad de Excelencia Mar\'{\i}a de Maeztu CEX2020-001058-M. {A.S-M. also acknowledges support from the RyC2021-032892-I grant funded by MCIN/AEI/10.13039/501100011033 and by the European Union `Next GenerationEU'/PRTR}. This work was partially supported by FAPESP (Funda\c{c}\~ao de Amparo \'a Pesquisa do Estado de S\~ao Paulo) under grant 2021/01183-8.

\vspace{3mm}
\facilities{ALMA}
\software{CASA}

\bibliography{biblio}{}
\bibliographystyle{aasjournal}

\vfil\eject

\appendix

\restartappendixnumbering

\section{1.3 mm continuum sources} \label{ap:sources}
\begin{figure}[ht]
\includegraphics[width=1\textwidth]{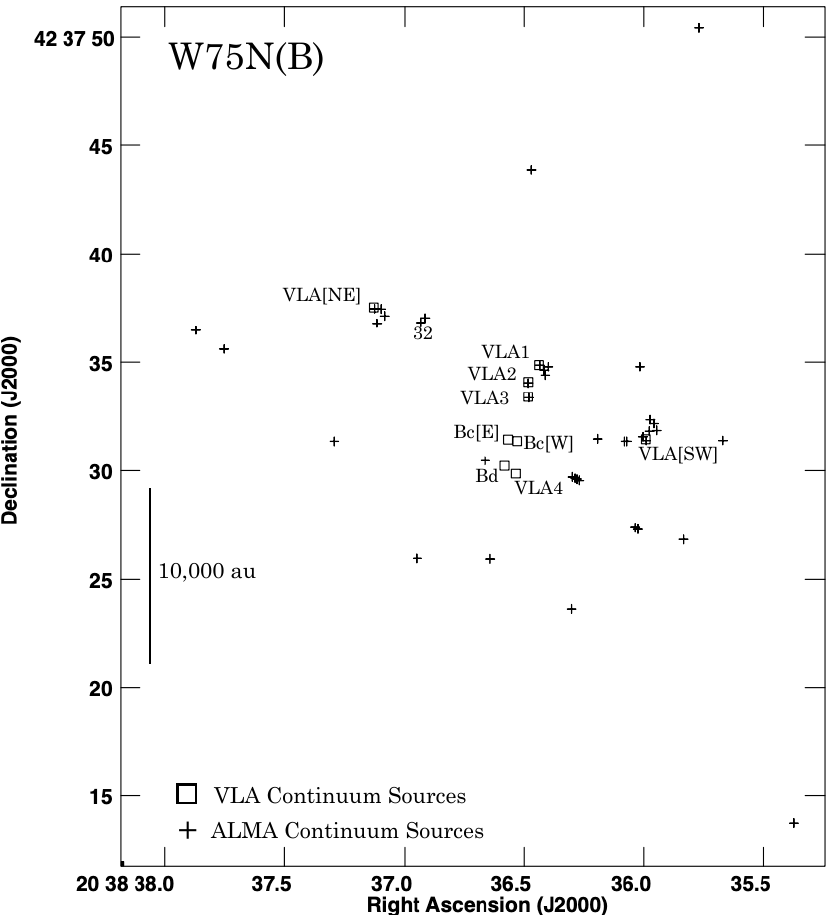}
\caption{Spatial distribution of the 1.3 mm continuum sources detected in a region of $\sim$30$''$ in size. Their positions (indicated here by crosses) and flux densities are listed in Table~\ref{tab:possources}. The powering source of the SiO bipolar jet also detected with our ALMA observations
(see \S\ref{ap:siojet} and Fig.~\ref{fig:siojet})
is labeled as \#32 (Table~\ref{tab:possources}). This source is located in between the two lobes of the SiO bipolar jet (Fig.~\ref{fig:siojet}). The positions of the VLA continuum sources \citep[][]{R-K20} are also labeled and indicated by squares. 
\label{fig:sources}}
\end{figure}

Table~\ref{tab:possources} lists the position of all the 1.3~mm continuum compact sources detected with ALMA in a region of $\sim$30$''$ ($\sim$39,000~au) centered on the massive protostar W75N(B)-VLA2 (see \S\ref{sec:dust}). Peak flux densities are primary beam corrected.
Their association (if any) with sources detected at other wavelengths are indicated in  the last column of Table~\ref{tab:possources}. Figure~\ref{fig:sources} shows the distribution of these mm continuum sources (indicated by crosses) together with the positions of the cm continuum sources detected with the VLA \citep[][]{R-K20}.
We are proposing, in particular, that one of these detected continuum sources (source ID 32 in Table~\ref{tab:possources}) is the powering source of the highly collimated bipolar jet that we have detected $\sim$6$''$ ($\sim$7,800 au) northeast of VLA2 (see Appendix~\ref{ap:siojet} and Fig.~\ref{fig:siojet}).

Dust mass estimates in the detected sources can be obtained from the 1.3 mm continuum emission assuming that it is optically thin and that the dust is isothermal within the beam size, through the equation
 \begin{equation}
M_{\rm dust} = \frac{S_\nu D^2}{\kappa_\nu B_\nu(T_{\rm dust})},
 \end{equation}
where $S_\nu$ is the observed flux density (Table~\ref{tab:possources}), D the source distance (1.3~kpc), $\kappa_\nu$ the dust opacity, and $B_\nu(T_{\rm dust})$ the Planck function at the dust temperature $T_{\rm dust}$. Adopting a dust opacity  $\kappa_\nu$(1.3 mm) = 0.899 cm$^2$ g$^{-1}$ \citep[][]{Ossenkopf94,D-R22}, we obtain 

\begin{equation} \label{eq:massdust}
M_{\rm dust} \simeq  1.4\times 10^{-4} M_{\odot} \left[\frac{S_{1.3~mm}}{mJy}\right] \times  \left[\frac{T_{\rm dust}}{50~K}\right]^{-1}.
 \end{equation}

These dust mass values should be taken with caution and as a first approximation, in particular because if the emission is not optically thin they represent lower limits. 
For sources with flux densities of $\sim$1~mJy (Table~\ref{tab:possources}), and assuming a gas-to-dust ratio of 100, total masses (dust+gas) of $\sim$1.5$\times$10$^{-2}$~M$_{\odot}$ ($\sim$15~M$_J$) are estimated for $T_{\rm dust}$ = 50~K, similar to the values obtained for protostellar disks of low-mass YSOs in massive star-forming regions, e.g., GGD 27 \citep[][]{Busquet19}, or Orion molecular clouds A and B \citep[][]{Tobin20}. This suggests that, in the W75N(B) massive star-forming region, there may also exist a significant population of low-mass YSOs (coexisting with massive protostars as VLA1, VLA2, and VLA3, and other possible intermediate-mass YSOs), having protostellar disks of size $\lesssim$170~au. 
Complementary observations at submm wavelengths will indeed be important in properly characterizing and distinguishing the nature of these sources as potential protostars.

\begin{deluxetable*}{ccccccc}[htb]

\tablecaption{ALMA 1.3 mm continuum sources detected in the W75N(B) field\tablenotemark{a}\label{tab:possources}}
\tablehead{\colhead{source ID} & \colhead{R.A. (J2000) } & \colhead{Dec. (J2000)} & \colhead{S$_{\nu}$ (mJy~beam$^{-1}$)\tablenotemark{b}}  &\colhead{$\Delta d$ (arcsec)\tablenotemark{c}} & \colhead{M$_{\rm J}$\tablenotemark{d}} & \colhead{Association\tablenotemark{e}}
}
\startdata
1    &  20 38 35.375  &  42 37 13.75  &  21.36  &  23.75 & 313 &  \nodata\\
2    &  20 38 35.671  &  42 37 31.39  &   2.45  &   9.39 & 36 &  \nodata\\
3    &  20 38 35.769  &  42 37 50.41  &   3.76  &  18.14 & 55 &  \nodata\\
4    &  20 38 35.837  &  42 37 26.82  &  0.94  &  10.21 & 14 &  \nodata\\
5    &  20 38 35.947  &  42 37 31.86  &   3.39  &   6.35 & 50 &  MM2 \\
6    &  20 38 35.959  &  42 37 32.17  &   8.24  &   6.13 & 121 &  MM2\\
7    &  20 38 35.974  &  42 37 32.35  &  3.87  &   5.91 & 57 &  MM2 \\
8    &  20 38 35.977  &  42 37 31.82  &   1.83  &   6.06 & 27 &  MM2 \\
9    &  20 38 35.991  &  42 37 31.39  &  12.68  &   6.09 & 186 &  MM2, VLA[SW]\\
10  &  20 38 36.004  &  42 37 31.57  &  6.66  &   5.89 & 98 &  MM2 \\
11  &  20 38 36.016  &  42 37 34.79  &  1.02  &   5.23 & 15 &  \nodata\\
12  &  20 38 36.025  &  42 37 27.29  &  0.75  &   8.49 & 11 &  \nodata\\
13  &  20 38 36.036  &  42 37 27.37  &  0.75  &   8.36 & 11 &  \nodata\\
14  &  20 38 36.070  &  42 37 31.35  &  0.75  &   5.35 & 11 &  MM2 \\
15  &  20 38 36.080  &  42 37 31.35  &  0.85  &   5.25 & 12 &  MM2 \\
16  &  20 38 36.194  &  42 37 31.47  &  1.88  &   4.15 & 28 &  MM2 \\
17  &  20 38 36.270  &  42 37 29.54  &  0.65  &   5.14 & 10 &  \nodata \\
18  &  20 38 36.279  &  42 37 29.60  &  0.78  &   5.04 & 11 &  \nodata\\
19  &  20 38 36.287  &  42 37 29.64  &  0.86  &   4.96 & 13 &  \nodata\\
20  &  20 38 36.298  &  42 37 29.70  &  1.05  &   4.86 & 15 &  \nodata\\
21  &  20 38 36.302  &  42 37 23.61  &  4.24  &  10.67 & 62 &  \nodata\\
22  &  20 38 36.401  &  42 37 34.79  &  2.16  &   1.17 & 32 &  MM1 \\
23  &  20 38 36.414  &  42 37 34.40  &  1.95  &   0.85 & 29 &  MM1 \\
24  &  20 38 36.418  &  42 37 34.63  &  1.60  &   0.92 & 23 &  MM1 \\
25  &  20 38 36.437  &  42 37 34.86  &  3.32  &   0.94 & 49 &  MM1, VLA1\\
26  &  20 38 36.471  &  42 37 43.87  &  0.39  &   9.78 & 6 &  MM[N] \\
27  &  20 38 36.482  &  42 37 33.36  &  86.93  &   0.73 & 1275 &  MM1, VLA3\\
28  &  20 38 36.484  &  42 37 34.04  &  20.75  &   0.05 & 51\tablenotemark{d} &  MM1, VLA2\\
29  &  20 38 36.641  &  42 37 25.93  &  0.99  &   8.34 & 15 &  \nodata\\
30  &  20 38 36.663  &  42 37 30.45  &  2.94  &   4.13 & 43 &  \nodata\\
31  &  20 38 36.912  &  42 37 37.03  &  0.55  &   5.55 & 8 &  \nodata\\
32  &  20 38 36.930  &  42 37 36.83  &  1.02  &   5.61 & 15 &  f \\
33  &  20 38 36.946  &  42 37 25.95  &  1.25  &   9.59 & 18 &  \nodata\\
34  &  20 38 37.083  &  42 37 37.13  &  1.87  &   7.26 & 27 &  MM3 \\
35  &  20 38 37.097  &  42 37 37.45  &  2.08   &   7.53 & 30 &  MM3 \\
36  &  20 38 37.114  &  42 37 36.79  &  1.26  &   7.44 & 18 &  MM3 \\
37  &  20 38 37.124  &  42 37 37.46  &  5.25  &   7.81 & 77 &  MM3, VLA[NE]\\
38  &  20 38 37.294  &  42 37 31.35  &  0.94  &   9.33 & 14 &  \nodata\\
39  &  20 38 37.752  &  42 37 35.60  &   12.46  &  14.05 & 183 &  \nodata\\
40  &  20 38 37.872  &  42 37 36.47  &  5.22  &  15.48 & 77 &  \nodata\\
\enddata
\tablenotetext{a}{Positions and peak intensities of the ALMA 1.3 mm continuum sources determined from the images with a beam size of $0.13''\times0.06''$ (PA = $-2^{\circ}$) and rms $\simeq$ 0.06 mJy~beam$^{-1}$ (see \S 3.1). The mm continuum sources are not resolved at FWHM, which implies physical sizes for the dust emission of $\lesssim$ 170 au. Positional uncertainties are $\sim$0.02$''$. Phase center is at $\alpha$(J2000) = 20$^h$38$^m$36.4860$^s$, $\delta$(J2000) = 42$^{\circ}$37$'$34.090$''$.} 
\tablenotetext{b}{Peak intensities corrected by primary beam.} 
\tablenotetext{c}{Distance from the source to the phase center of the observations.}
\tablenotetext{d}{Total mass (dust+gas) from Eq.~\ref{eq:massdust}, assuming $T_{\rm dust}$ = 50~K (see \S\ref{ap:sources}), excepting for source ID 28 (VLA2), where we assume $T_{\rm dust}$ = 300~K (see \S\ref{sec:toroid}).}
\tablenotetext{e}{The association with the mm cores MM1, MM2, MM3, MM[N], and/or the VLA sources VLA[SW], VLA1, VLA2, VLA3, VLA[NE] \citep[][]{R-K20} is indicated.}
\tablenotetext{f}{Source proposed to power the highly-collimated SiO bipolar outflow detected $\sim$6$''$ northeast of VLA2 (see \S\ref{ap:siojet} and Fig.~\ref{fig:siojet}).} 
\end{deluxetable*}

\vfil\eject

\restartappendixnumbering

\section{H$_2$CO and SiO absorption lines toward VLA3}\label{ap:absorption}

\begin{figure}[ht]
\includegraphics[width=1\textwidth]{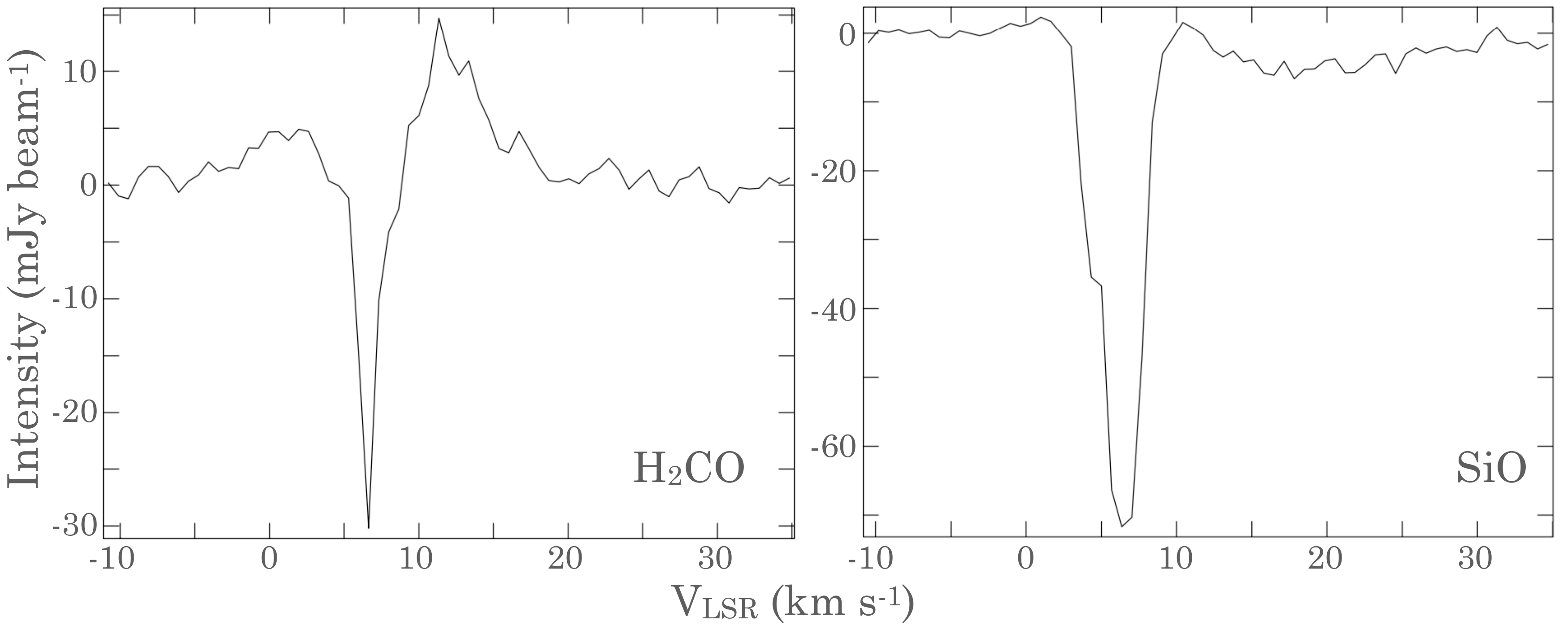}
\caption{H$_2$CO (left panel) and SiO (right panel) lines seen in absorption toward the continuum source VLA3.
\label{fig:absorption}}
\end{figure}

{ H$_2$CO and SiO lines are seen in absorption against the continuum source VLA3. In Fig. \ref{fig:absorption} we present the spectra at the location of this source. Both show an absorption feature centered at $V_{\rm LSR}\simeq 6.4$ km s$^{-1}$.}

\vfil\eject

\section{Highly collimated bipolar SiO jet} \label{ap:siojet}

\begin{figure}[ht]
\includegraphics[width=1\textwidth]{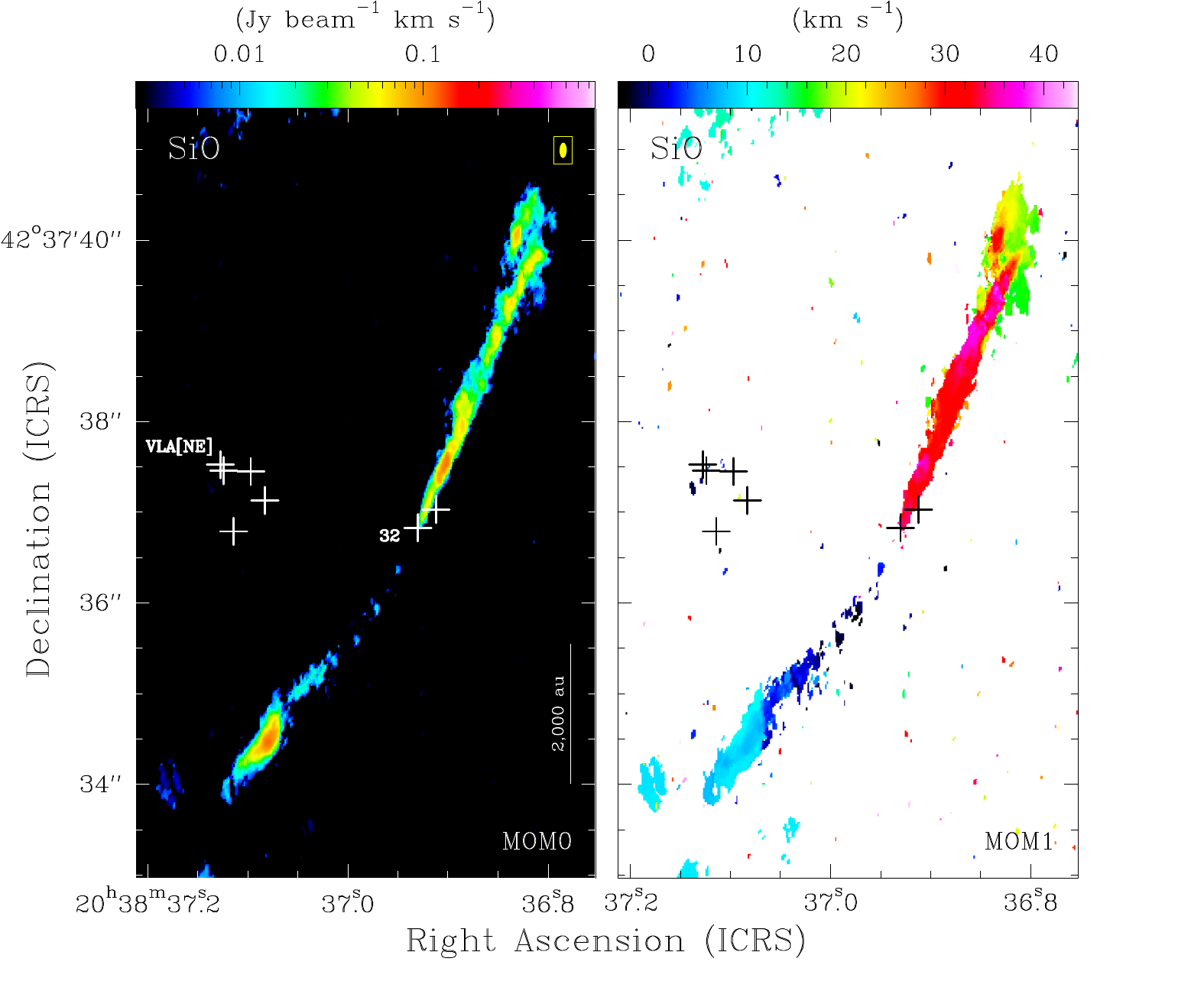}
\caption{{\it Left panel:} Integrated intensity image (moment of order 0) of the 
SiO bipolar jet found $\sim5''$ northeast of VLA2 in the velocity range $V_{\rm LSR} \simeq -3$ to +43~km s$^{-1}$.
The position of the 1.3 mm continuum source proposed to power this SiO jet  (\S\ref{ap:siojet}, Table~\ref{tab:possources}) is indicated by a cross and labeled as number 32. The nearby radio continuum source VLA[NE] \citep[][]{R-K20} is also labeled. Other 1.3 mm continuum sources found nearby of the SiO jet are also indicated by crosses. The synthesized beam is marked as a filled yellow ellipse at the upper right corner. {\it Right panel:} Velocity field image of the SiO jet (moment of order 1). 
\label{fig:siojet}}
\end{figure}

We report here the detection of an extremely highly 
collimated bipolar SiO jet  located $\sim$5$''$ ($\sim$6,500~au) northeast of the three massive protostars (Fig.~\ref{fig:siojet}). This bipolar  jet extends in the southeast-northwest direction, over an angular scale of $\sim 8''$ ($\sim 10,400$ au) and velocities in the range $V_{\rm LSR} \simeq -3$ to +43 km s$^{-1}$. From the detected compact mm continuum sources in the W75N(B) region (\S\ref{sec:dust}), we identify one of them (source ID 32 in Table~\ref{tab:possources}) that is well-centered  between the two jet lobes (Fig.~\ref{fig:siojet}), as the most probable powering source, most likely a low-mass protostar. Since it is located in a region outside the main focus of our observations on VLA2, we will present and discuss the main properties of this remarkable bipolar SiO jet in a forthcoming paper.

\restartappendixnumbering

\vfil\eject

\end{document}